\documentclass[aps,prc,twocolumn,preprintnumbers,groupedaddress,superscriptaddress]{revtex4}
\usepackage{graphicx}
\usepackage{color}

\newcommand{\pT}{$p_T$ }
\newcommand{\mT} {$m_T$ }

\newcommand{\sNN}{$\sqrt{s_{_\mathrm{NN}}}$ }
\newcommand{\s}{$\sqrt{s}$ }
\newcommand{\pp}{$p$+$p$ }

\newcommand{\auau}{Au+Au }

\usepackage{hyperref}
\begin{document}

\preprint{}

\title{Onset of radial flow in $p$+$p$ collisions}




\author{Kun Jiang}
\affiliation{Department of Modern Physics, University of Science and Technology of China, Hefei
230026, China}
\author{Yinying Zhu}
\affiliation{Department of Modern Physics, University of Science and Technology of China, Hefei
230026, China}
\author{Weitao Liu}
\altaffiliation{Now at Department of Physics and Astronomy, University of Bonn, Bonn, Germany.}
\affiliation{Department of Modern Physics, University of Science and Technology of China, Hefei
230026, China}

\author{Hongfang Chen}
\affiliation{Department of Modern Physics, University of Science and Technology of China, Hefei
230026, China}
\author{Cheng Li}
\affiliation{Department of Modern Physics, University of Science and Technology of China, Hefei
230026, China}
\author{Lijuan Ruan}
\affiliation{Physics Department, Brookhaven National Laboratory, Upton, NY11973, USA}
\author{Zebo Tang}
\email{zbtang@ustc.edu.cn}

\affiliation{Department of Modern Physics, University of Science and Technology of China, Hefei
230026, China}
\author{Zhangbu Xu}
\affiliation{Physics Department, Brookhaven National Laboratory, Upton, NY11973, USA}


\date{\today}

\begin{abstract}
It has been debated for decades whether hadrons emerging from $p$+$p$ collisions exhibit collective expansion. The signal of the collective motion in $p$+$p$ collisions is not as clear/clean as in heavy-ion collisions because of the low multiplicity and large fluctuation in $p$+$p$ collisions. Tsallis Blast-Wave (TBW) model is a thermodynamic approach, introduced to handle the overwhelming correlation and fluctuation in the hadronic processes.  We have systematically studied the identified particle spectra in $p$+$p$ collisions from RHIC to LHC using TBW and found no appreciable radial flow in $p$+$p$ collisions below $\sqrt{s}=900$~GeV. At LHC higher energy of 7~TeV in $p$+$p$ collisions, the radial flow velocity achieves an average value of $\langle \beta \rangle = 0.320\pm0.005$. This flow velocity is comparable to that in peripheral (40-60\%) Au+Au collisions at RHIC. Breaking of the identified particle spectra $m_T$ scaling was also observed at LHC from a model independent test.

\end{abstract}


\maketitle



\section{Introduction}
The searches for a Quark-Gluon Plasma (QGP) have been conducted in hadron collisions in all collision energies and species. Many have argued that some features observed in \pp collisions at high multiplicity and/or high energy resemble a QGP. The most acclaimed evidence has been the observation of a collective expansion~\cite{Alexopoulos:1993wt,PhysRevLett.67.1519,arXiv1010.0964}. However, what constitutes a collective expansion when the particles reach our detectors are free streaming by nature? While it is seemingly trivial to argue that flow is a mass effect and therefore a systematic enhancement of heavier particles at higher momentum~\cite{BraunMunzinger:1994xr,fuqiangLongPID,Chatrchyan:2012qb,cms-pas-fsq-12-014} would be a signature of flow, large fluctuation in temperature and/or the creation of mini-jets in semi-hard processes can produce similar qualitative features~\cite{Wang:1988bw,Wang:1991us,Wang:1991vx}. Hydrodynamic simulation with small viscous correction has been successful in interpreting many phenomena observed in heavy-ion collisions. However, its applicability to \pp collisions with large fluctuation and viscosity is not obvious.

With increasing colliding energy in \pp collisions, two possible phenomena emerge: color glass condensate (CGC) and holographic pomeron model mathematical equivalence to black hole radiation in 5+5 dimensions~\cite{Shuryak:2013ke}. At LHC energies, model incorporating CGC~\cite{Dumitru:2010iy} correctly describes the CMS data on di-hadron correlation~\cite{Khachatryan:2010gv} without flow while the argument from black hole radiation predicts large radial flow in \pp collisions at high multiplicity~\cite{Shuryak:2013ke, Shuryak:2013sra}. Recently, on-going debates focus on whether hydrodynamics are applicable to small system when such a system has large shear viscous effect by design. It is therefore problematic for the elliptic flow to be quantitatively interpreted in a hydrodynamic evolution for \pp collisions.  However, radial flow is expected to be less affected by the viscous correction. Anisotropic flow is by definition a relative quantity while radial flow velocity is an absolute velocity. Extracting this radial velocity has been at qualitative level and is model dependent in both \pp and A+A collisions. The main reason of the failure is that radial flow is not the dominant feature in identified particle spectra in \pp collisions and to a progressively lesser degree in A+A collisions.

Although it is known that fragmentation from hard processes and hadronization in QCD contribute significantly to the particle production at low momentum, it has been a subject of investigation to find an elegant approach to incorporate these phenomena in a thermodynamic or statistical approach. The framework allows application of hydrodynamic-inspired blast-wave model~\cite{blastWave} to extract flow velocity while being able to correctly fit the available data with very good $\chi^2$ per degree of freedom (ndf) in a large transverse momentum range. This is the philosophy presented in this paper. We use a non-extensive thermodynamic model, Tsallis statistics~\cite{q-entropy}, to describe the particle production from QCD hadronization including jet contribution. We incorporated it into the blast-wave expansion to fit data and extract flow velocity and other thermodynamic parameters~\cite{Tang:2008ud, Shao:2009mu, Tang:2011xq}.
The model can be vetted by its simplicity in interpreting physics phenomena and by achieving best $\chi^2$ description of data. We emphasize that this is not to replace the more fundamental QCD theory or hydrodynamic simulation. On the contrary, the method resembles an ``experimental" approach to extract physical quantities from data, which can then be concisely used to compare with elaborated theories.

This paper is organized as follows: we present the analysis method of all the identified particle spectra in \pp collisions at $\sqrt{s}=200$, 540, 900 and 7000 GeV. A two-particle correlation function is also introduced in this paper based on TBW model. The results from the TBW fits to the data are presented in subsequent section. The result provides an onset of beam energy where radial flow has been developed in minimum-bias \pp collisions. At the end, possible improvement and more data collection and analyses are discussed.

\section{Analysis Method}
Similar to what presented in the literatures~\cite{blastWave, starWhitePaper, fuqiangLongPID, Retiere_Lisa_PRC70, Tang:2008ud, Shao:2009mu, Tang:2011xq}, we have used the TBW model to extract thermodynamic and hydrodynamic quantities from data. The single-particle spectrum can be written as
\begin{eqnarray}\label{eq:spectra}
\frac{{d}^{2}N}{2\pi m_{_T} dm_{_T}dy}|_{y=0} = A \int_{-y_b}^{+y_b}
e^{\sqrt{y_b^{2}-y_{s}^{2}}} m_{_T}\cosh \left( y_s \right)dy_s
\nonumber \\
\times\int_{0}^{R}rdr \int_{-\pi}^{\pi} \left [ 1+\frac{q-1}{T} E_T \right ]^{-1/(q-1)}d\phi.
\end{eqnarray}
Where
\begin{equation}\label{eq:mT}
m_T = \sqrt{p_T^2+m^2},
\end{equation}

\begin{equation}\label{eq:yb}
y_b = \ln \left( \sqrt{s_{\mathrm{NN}}}/m_N \right),
\end{equation}

\begin{equation}\label{eq:ET}
E_T=m_T\cosh(y_s)\cosh(\rho) -p_T\sinh(\rho)\cos(\phi).
\end{equation}
$A$ is a normalization factor, $m$ is the mass of the particle, $m_N$ is the mass of the colliding nucleon, $y_s$ is the rapidity of the emitting source, $y_b$ is the beam rapidity and $\phi$ is the azimuthal angle between the flow velocity and the emitted particle velocity in the rest frame of the emitting source. The emitting source are boosted with the boost angle
\begin{equation}\label{eq:rho}
\rho = \tanh^{-1} \left[ \beta_S \left( \frac{r}{R}\right)^{n} \right].
\end{equation}
Where $r$ is the radius of the emitting source, $\beta_S$ is the velocity of the source at the outermost radius ($r=R$), $n$ (=1) determines the source velocity profile.

One of the significant advantages of TBW in comparison to the Boltzmann-Gibbs Blast-Wave (BGBW) is the capability of describing a system with large fluctuation and correlation, which is the case of $p$+$p$ collisions. Based on the nonextensive Tsallis statistics, the temperature distribution of the nonequilibrium system is characterized by the parameters $q$ and $T$, where $T$ is related to the average of the inverse temperature and the nonextensivity parameter $q$ can be interpreted as its fluctuation~\cite{Wilk:1999dr,Wilk:2008ue,Biro:2008hz}. The Tsallis distribution converges to Boltzmann-Gibbs distribution when $q$ tends to unity. When $q-1$ is small, the TBW approach is not different from many treatments on dissipative hydrodynamics with a small perturbation around Boltzmann distribution~\cite{Beck:2001ts, Kodama:2005pp, Osada:2008sw, Biro:2011bq}. In TBW, the free parameters required to predict the \pT spectra of a given particle species are $\beta_S$, $T$, $q$ and $A$. If only the shape is concerned, the normalization factor $A$ is not needed.

In recent theory development, the correlations originated from initial gluon scattering could be enhanced by the radial pressure from bulk flow~\cite{Bozek:2010pb,Werner:2010ss}. K. Dusling and R. Venugopalan \cite{Dusling:2012iga} presented a schematic description of the enhancement. It has been argued that significant radial flow has been ruled out by the di-hadron correlation from CMS~\cite{Khachatryan:2010gv}. It is therefore imperative to study the correlation effect in the present of radial flow in \pp collisions. To implement such effect in TBW, we have introduced an anisotropic emission of particles from the source to account the particles produced from the initial correlated gluon fragmentation. The anisotropic emission is described as
\begin{equation}
\frac{dN}{d\phi} \propto 1+2 p_{_2} \cos(2\phi).
\end{equation}
where $\phi$ represents the angle between the individual emitted particle and the back-to-back axis. The TBW formula becomes
\begin{eqnarray}\label{eq:spectra2}
\frac{{d}^{2}N}{2\pi m_{_T} dm_{_T}dy}|_{y=0} = A \int_{-y_b}^{+y_b}
e^{\sqrt{y_b^{2}-y_{s}^{2}}} m_{_T}\cosh \left( y_s \right)dy_s
\nonumber \\
\times\int_{0}^{R}rdr \int_{-\pi}^{\pi} \left[1+2 p_{_2} \cos(2\phi)\right]\nonumber \\
\times\left[ 1+\frac{q-1}{T} E_T \right ]^{-1/(q-1)}d\phi.
\end{eqnarray}
The azimuthal anisotropy coefficient $c_2$ can be obtained through
\begin{equation}\label{eq:v2}
c_2(p_{_T})=\langle \cos(2\phi)\rangle.
\end{equation}
The correlated distribution is on top of a large isotropic underlying event background. Taking this contribution into account, $c_2$ becomes
\begin{equation}\label{eq:v2-s2}
c_2(p_{_T})=s_{_2} \langle \cos(2\phi)\rangle,
\end{equation}
where $s_{_2}$ depicts the fraction of the anisotropic emitting source ($0 \leq s_{_2} \leq 1$). The di-hadron correlation can be obtained from the $c_2$ of hadrons through
\begin{equation}
\frac{dN^{\textrm{Assoc}}}{N^{{\textrm{Trig}}}d(\Delta\phi)} = \frac{N^{{\textrm{Assoc}}}}{2\pi}
\left[ 1+ c_2^{\textrm{Trig}} c_2^{\textrm{Assoc}} \cos(2\Delta\phi) \right].
\end{equation}

It is important to note that this procedure is different from the implementations of elliptic flow in the blast-wave model (e.g. \cite{Retiere_Lisa_PRC70,Tang:2011xq}). Here we focus on the collimation of the initial azimuthal correlation by radial flow.

\begin{figure*}[]
\centering
\includegraphics[width=0.75\textwidth]{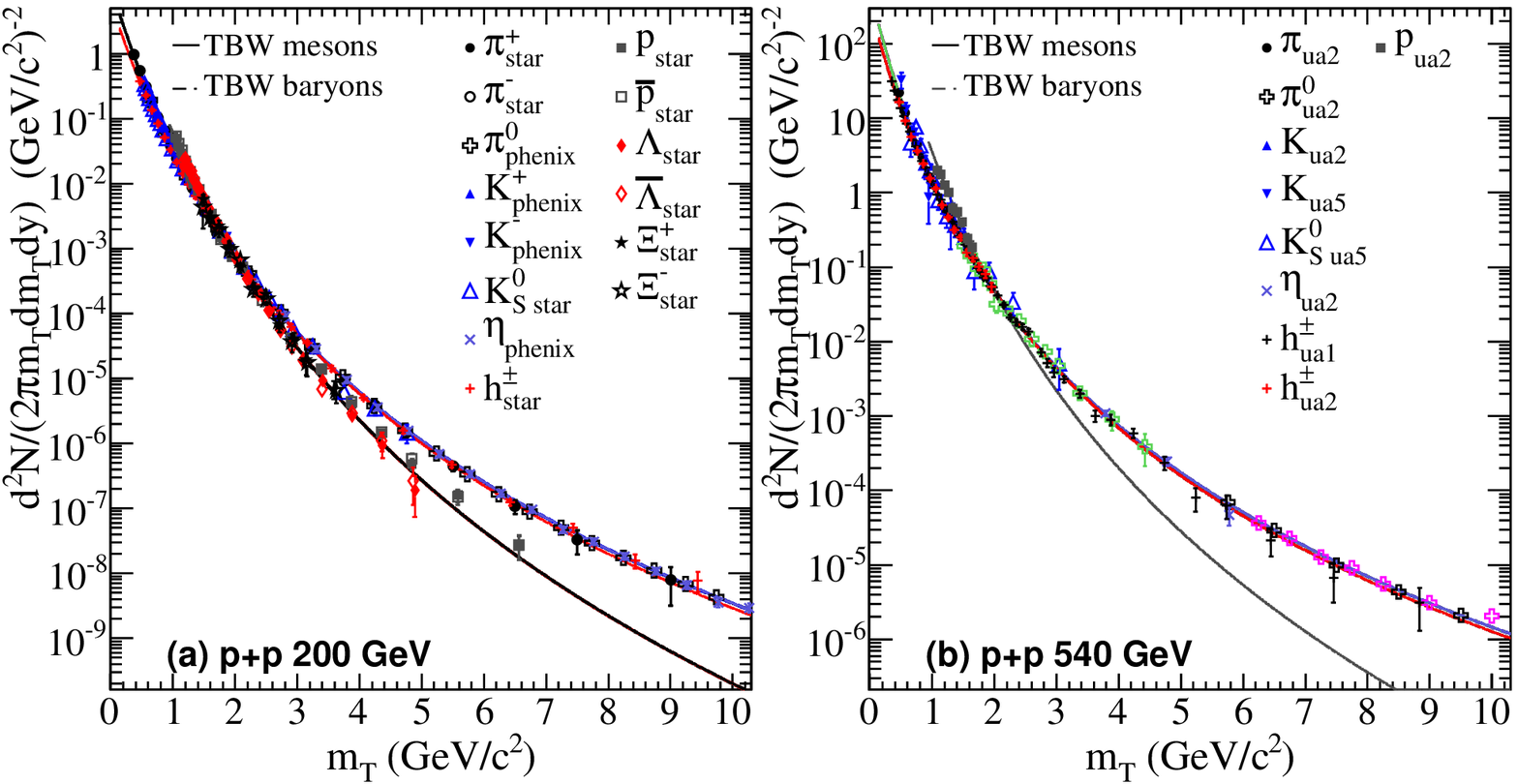}
\includegraphics[width=0.75\textwidth]{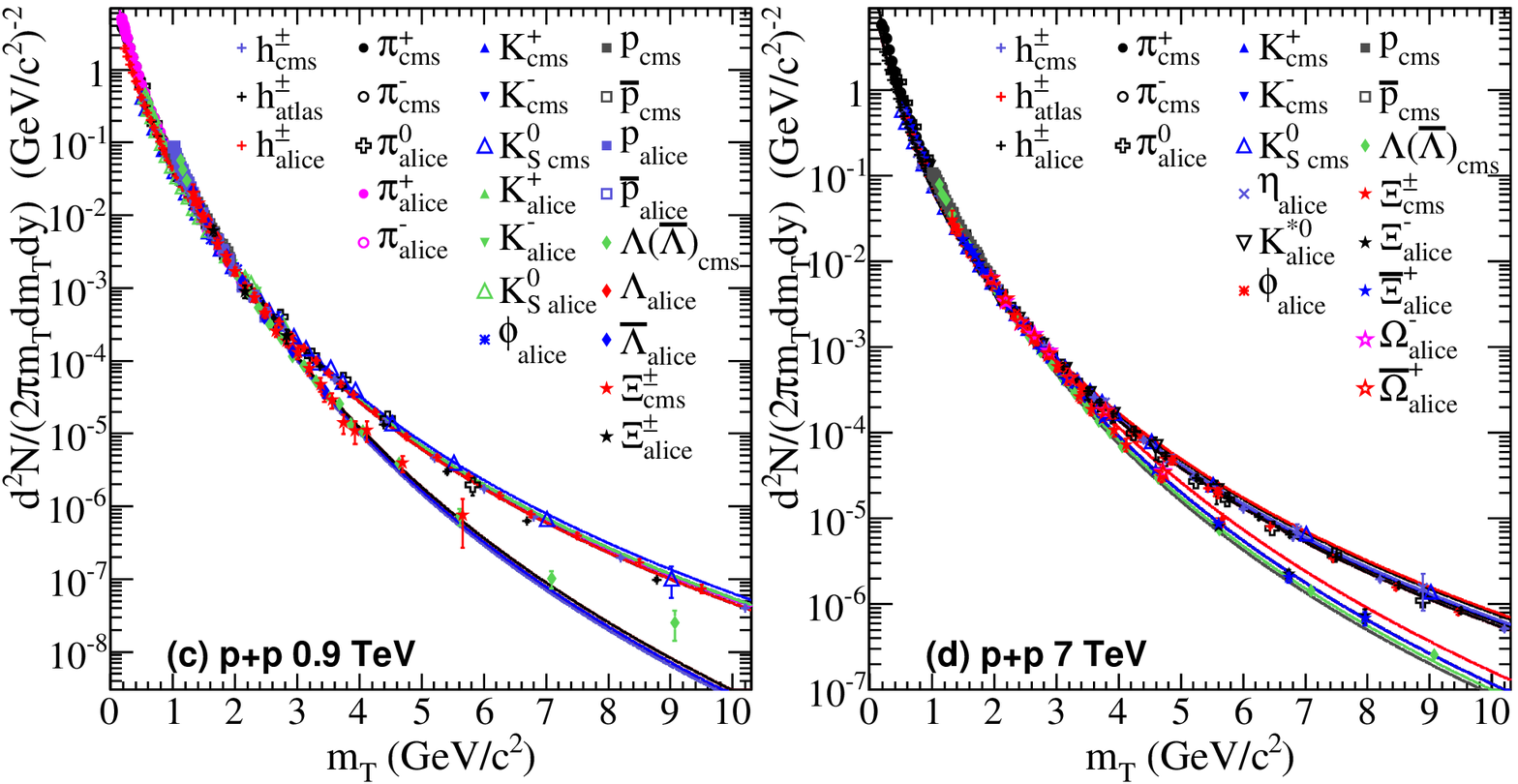}
\caption{Identified particle \mT spectra in \pp collisions at \s = 200~GeV (a), 540~GeV (b), 0.9~TeV (c) and 7~TeV (d). The symbols represent experimental measurements and the curves represent TBW fit results. At each energy, all of the \mT spectra are rescaled to have the same value at $m_T=2~\textrm{GeV}/c^2$ as $\pi^+$. The references of the experimental measurement are summarized in Tab.~\ref{tab:dataRef}.}
\label{fig:spectra}
\end{figure*}

STAR and PHENIX at RHIC, UA1, UA2 and UA5 at Sp$\bar{\textrm{p}}$S, E735 at FermiLab, and CMS and ALICE at LHC have published a comprehensive collection of identified particle spectra in \pp collisions at 200, 540, 900 GeV and 7 TeV. Table~\ref{tab:dataRef} lists the available data from each reference from the collaborations. They are all from minimum bias (non-single-diffractive or non-diffractive) events. The particle \pT spectrum from different type of minimum bias events only differs by an overall normalization factor. The shape is the same.

\begin{table}[]
\caption{Summary of the data references.}\label{tab:dataRef}
\begin{tabular}{lcccccccc}
\hline
 &$\pi^{\pm}$, $K^{\pm}$,$p$ & $\pi^0$, $\eta$ & $K_S^0$, $\Lambda$, $\Xi^\pm$, $\Omega$   &   $K^{\star0}$, $\phi$&$h^{\pm}$\\
 \hline
 STAR & \cite{Adams:2006nd} &  & \cite{Abelev:2006cs} & & \cite{Adams:2003kv}&\\
 PHENIX & \cite{Adler:2006xd}& \cite{Adare:2007dg,Adare:2010cy} & & & &\\
 UA1 &                            &                                           &                      & & \cite{Arnison:1982ed}   &\\
 UA2 & \cite{Banner:1983jq}       & \cite{banner1985inclusive,Banner:1982zw}  &  \cite{Banner:1983jq}                     &                         & &\\
 UA5 & \cite{Alner:1985ra}        &                                           & \cite{Alner:1985ra}  &                         & &\\
 E735& \cite{Alexopoulos:1993wt}  &                                           &                      &                         & &\\

 CMS  & \cite{Chatrchyan:2012qb}   &   & \cite{Khachatryan:2011tm}  &
 & \cite{Chatrchyan:2011av}  &\\
 ALICE & \cite{Aamodt:2011zj} & \cite{Abelev:2012cn}  & \cite{Abelev:2012jp,Aamodt:2011zza} & \cite{Abelev:2012hy} & \cite{Abelev:2013ala} &\\
 ATLAS &  &   &  & & \cite{Aad:2010ac} &\\
\hline
\end{tabular}
\end{table}

\begin{table*}
\caption{Summary of the parameters.}\label{tab:para}
\centering
\begin{tabular}{lccccc}
  \hline
   \s & $\langle \beta \rangle$   & T (MeV) & $q_M-1$ & $q_B-1$ & $\chi^2$/ndf \\
  \hline
   ~~7~~ TeV & $0.320\pm0.005$ & $70.3\pm0.8$ & $0.1314\pm0.0003$ & $0.1035\pm0.0008$ & $490/431$ \\
900 GeV & $0.264\pm0.005$ & $74.6\pm0.5$ & $0.1127\pm0.0003$ & $0.0827\pm0.0008$ & $545/501$ \\
540 GeV & $0.000^{+0.105}_{-0.000}$ & $81.8\pm0.6$ & $0.1158\pm0.0007$ & $0.0841\pm0.0036$ & $205/168$ \\
  200 GeV & $0.000^{+0.124}_{-0.000}$ & $92.3\pm2.7$ & $0.0946\pm0.0006$ & $0.0743\pm0.0015$ & $268/268$ \\
  \hline
\end{tabular}
\end{table*}

\begin{figure*}[t]
\centering
\includegraphics[width=0.75\textwidth]{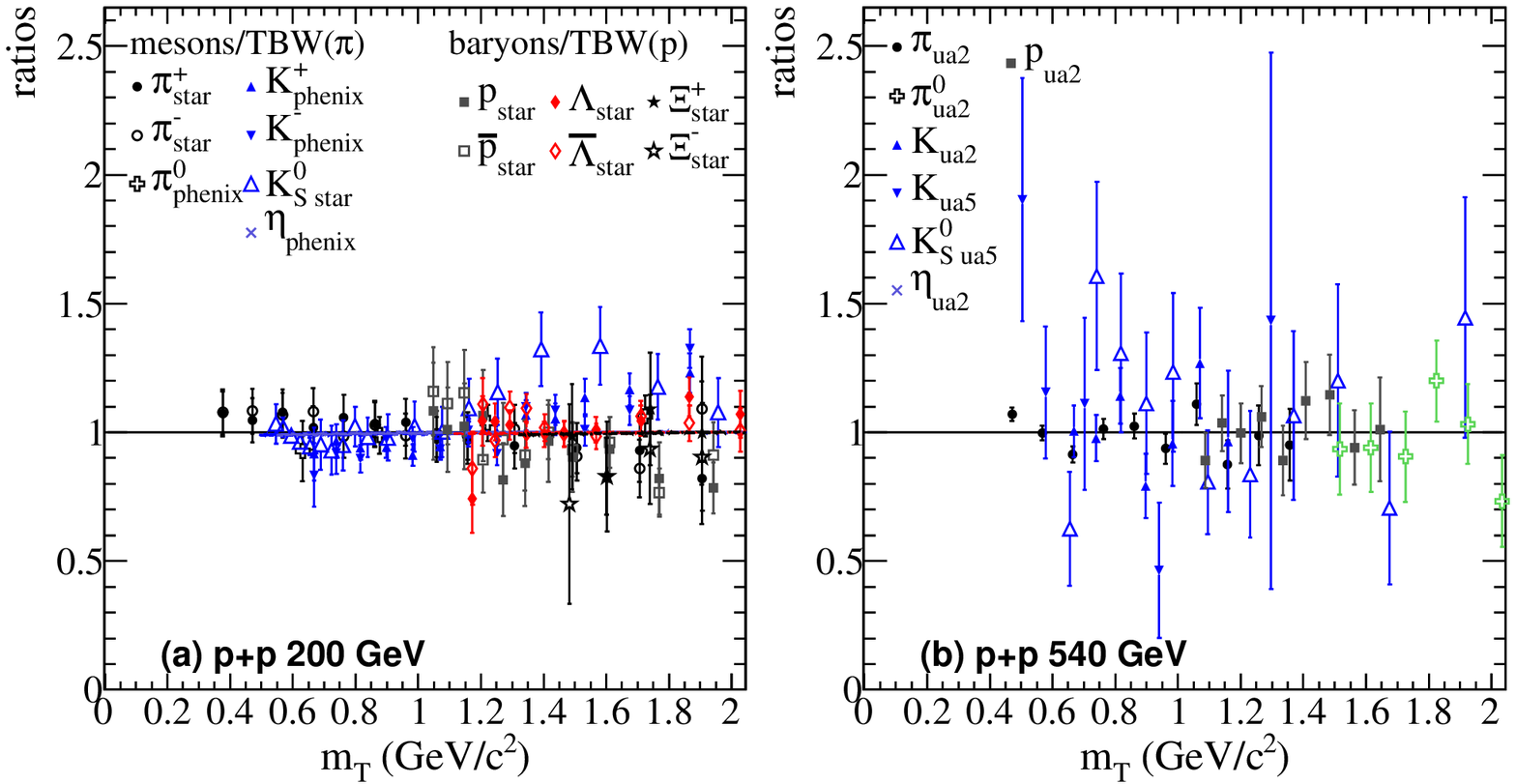}
\includegraphics[width=0.75\textwidth]{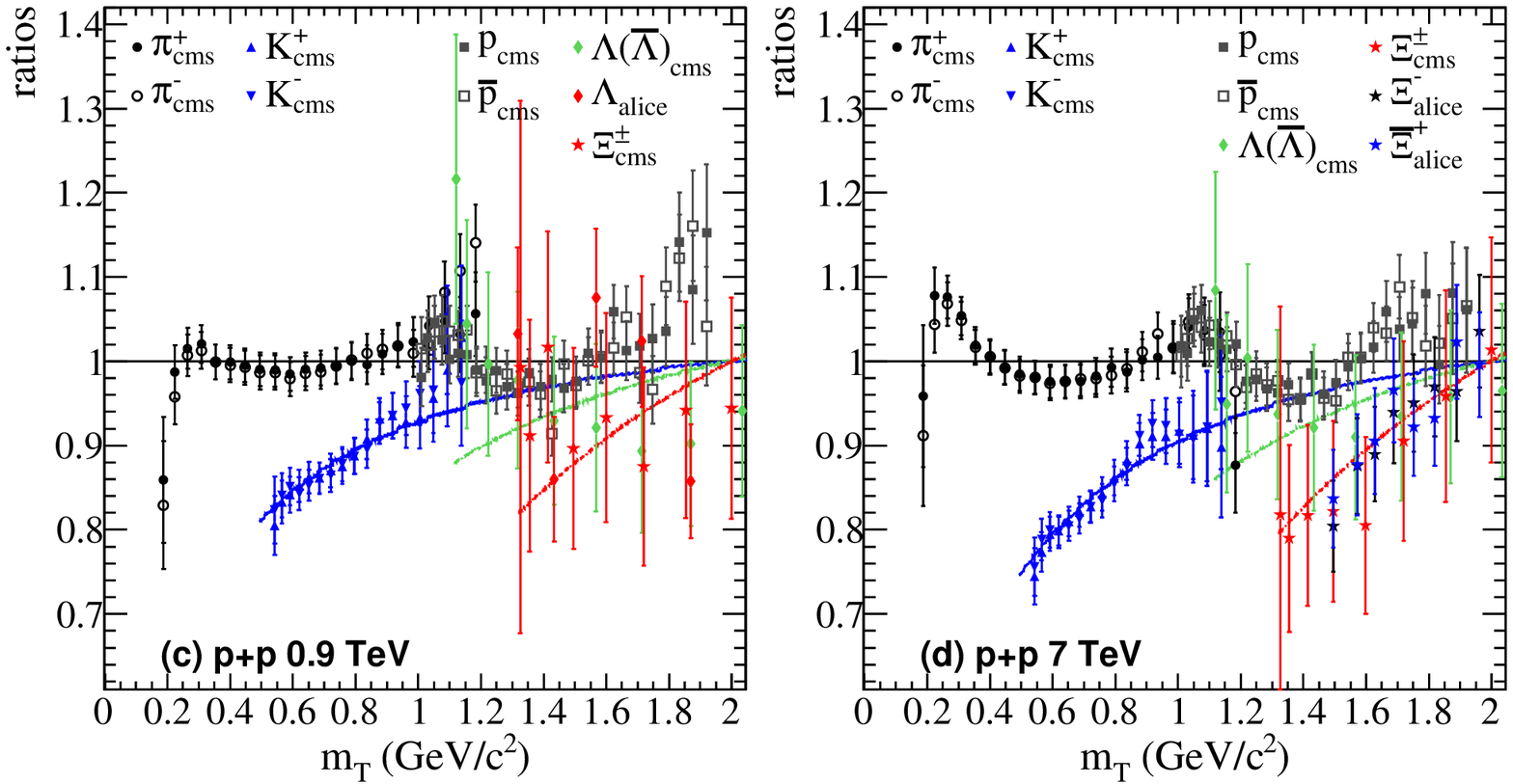}
\caption{\mT scaling behavior of the identified particle spectra in \pp collisions at \s=200~GeV (a), 540~GeV (b), 0.9~TeV (c) and 7~TeV (d). For mesons (baryons), the data points represent the ratio of rescaled \mT spectra shown in Fig.~\ref{fig:spectra} to the corresponding TBW curve of $\pi^+$ ($p$). }
\label{fig:ratio}
\end{figure*}

Figure \ref{fig:spectra}(c) shows the \mT spectra of $\pi^\pm$, $\pi^0$, $K^\pm$, $K_S^0$, $p$, $\bar{p}$, $\Lambda$($\bar{\Lambda}$), $\Xi^{\pm}$ and inclusive charged hadrons in \pp collisions at \s = 900 GeV. The \pT spectra of these particles are fit simultaneously with TBW (Eq.~\ref{eq:spectra}). There fit parameters and the best $\chi^2$ per fitting degree of freedom ($\textrm{ndf}$) are listed in Tab.~\ref{tab:para}. The parameters $\langle \beta \rangle=2\beta_S/3$ and $T$ are common to all of the particle species. The parameters $q_M$ and $q_B$ are common to all of the mesons and baryons, respectively. In addition to these 4 common parameters, each particle species has its own normalization factor $A$. The fit function for the inclusive charged hadron is the sum of that for $\pi^\pm$, $K^\pm$, $p$ and $\bar{p}$. We performed a least-$\chi^2$ fit of the 24 \pT spectra simultaneously with the TBW functions controlled by the $4+24$ parameters. Then the \pT spectra are converted to \mT spectra and rescaled to have the same value at $m_T=2~\textrm{GeV}/c^2$ as $\pi^+$, as shown in Fig.~\ref{fig:spectra}(c). The pion mass is applied for inclusive charged hadron when we do the \pT to \mT spectra conversion. The data and fit curve have the same rescale factor. Figure ~\ref{fig:spectra}(d) and \ref{fig:spectra}(a, b) show the rescaled identified hadron and inclusive charged hadron \mT spectra in \pp collisions at \s = 7~TeV, 200 and 540 GeV. The TBW fit curves are shown for all the particles as well.

In all the energies, all the spectra display power-law behavior at high \mT with grouping of baryons and mesons. The TBW describes the shape of the \mT spectra of more than 10 particles over a broad \mT range (0-10 GeV/$c^2$) at each energy, with only 4 quantities, as listed in Tab.~\ref{tab:para}. The quality of the fits are very good, the ratio of $\chi^2/\textrm{ndf}$ are between 1.00 and 1.22. At LHC energy, the radial flow velocity achieved an average value of $\langle \beta \rangle = 0.320 \pm 0.005$ and $0.264 \pm 0.005$ in \pp collisions at 7~TeV and 900~GeV, respectively. The velocity is comparable to that in peripheral (40-60\%) \auau collisions at \sNN = 200~GeV at RHIC ($0.282\pm0.017$~\cite{Tang:2008ud}). While at \s = 540~GeV and 200~GeV, the velocity in \pp collisions is consistent with zero ($\langle \beta \rangle = 0.000^{+0.105}_{-0.000}$ and $0.000^{+0.124}_{-0.000}$, respectively). The parameter $q$ is found to increase with increasing beam energy, and it is significantly higher for meson than for baryon at all of the energies. $T$ shows a reverse dependence on beam energy. The experimental observation of meson/baryon grouping~\cite{ppSplit} is described by the TBW model with two different $q$ parameters, but the extract physics implication is to be understood.

The \mT spectra of identified hadrons was found to have a universal behavior in high-energy \pp collisions, as known as \mT scaling. Equation \ref{eq:spectra}-\ref{eq:rho} show that if there is a non-zero radial flow, the shape of the \mT spectra depends not only on $m_T$, but also on $p_T$. This means the \mT scaling will be broken if there is a non-zero radial flow. To have a closer look at the effects on the \mT spectra induced by the non-zero radial flow, we tested the \mT scaling behavior of the identified particle spectra in \pp collisions as shown in Fig.~\ref{fig:ratio}. To illustrate the effect in linear scale, all of the data points and fit curves (shown in Fig.~\ref{fig:spectra}) for mesons are divided by the fit curve of $\pi^+$, those for baryons are divided by the fit curve of $p$. In \pp collisions at 900 GeV, as shown in Fig.~\ref{fig:ratio}(c), the ratio for $K^\pm$ is significantly below unity, and decreases as decreasing $m_T$. The $\Xi^\pm$ data points are also systematically below unity despite the large uncertainties. At higher beam energy, the deviation from the \mT scaling for $K^\pm$ and $\Xi^\pm$ is larger and more clear. It is clearly seen that the \mT scaling of identified particle spectra in \pp collisions is broken at beam energy above 900~GeV. This breaking can be described by the TBW with non-zero radial flow velocity very well. At lower energy, all the spectra still follow the \mT scaling, as shown in Fig.~\ref{fig:ratio} (a, b).

\begin{figure}[t]
\centering
\includegraphics[width=0.95\columnwidth]{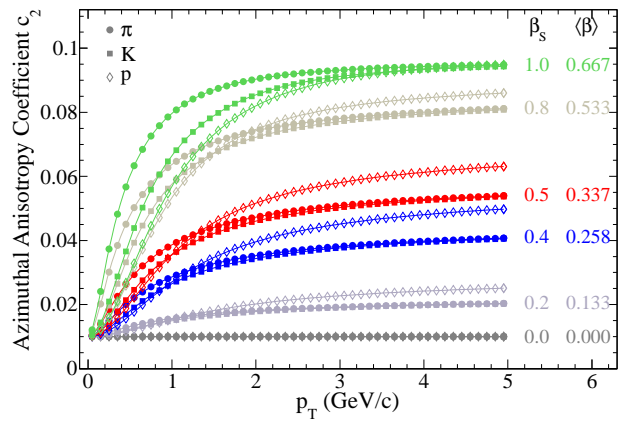}
\caption{The azimuthal anisotropy coefficient $c_2$ versus \pT for pion (solid circles), kaon (solid squares) and proton (open diamonds), illustrating the radial flow effect in Eq.~\ref{eq:v2-s2}. Both $p_2$ and $s_2$ are assumed to be 10\%. $T$, $q_M$ and $q_B$ are fixed to the values extracted from 7~TeV data (67.9~MeV, 1.1315 and 1.1009, respectively). The points with different colors are corresponding to different radial flow velocities.}
\label{fig:v2}
\end{figure}

To illustrate how radial flow boosts the particle collinear emission and enhances a pre-existing angular correlation, we assume that there is an existing correlation originating from initial condition and manifesting itself as anisotropic emission from its source at rest with $p_2$, and only a fraction of all emission source ($s_2$) possesses this characteristics and been driven by the later stage bulk radial flow. The scenarios are independent of hadron \pT and source location, and only serve for illustration purpose and are likely not realistic. Figure~\ref{fig:v2} shows the azimuthal anisotropy coefficient $c_2$ as a function of \pT for pion, kaon and proton, predicted by TBW according to Eq.~\ref{eq:v2-s2}. The parameters $p_2$ and $s_2$ are assumed to be 10\%. This means the fraction of initial anisotropic source is 10\%, and the particles emitted from the anisotropic source has $c_2=10\%$. The parameters $T$, $q_M$ and $q_B$ are fixed to the values obtained from the fit to the \pT spectra at 7 TeV. The radial flow velocity $\beta_S$ varies from $0.0$ to $1.0$ (from bottom to top). When there is no radial flow ($\beta_S=0$), $c_2$ is a constant of $10\%\times10\%=1\%$. Once there is a non-zero radial flow, $c_2$ is enhanced depending of the magnitude of radial flow velocity and $p_T$. It increases rapidly at low-\pT ($p_T\lesssim1~\textrm{GeV}/c$) and then tend to saturate. The mass ordering at low-\pT and baryon/meson grouping at intermediate- and high-\pT range is reproduced. In the whole \pT range, the predicted $c_2$ increases with increasing radial flow velocity. For the radial flow velocity of what we extracted from the 7~TeV data ($\langle\beta\rangle=0.320$), the saturated $c_2$ at $p_T\gtrsim2~\textrm{GeV}/c$ is predicted to be about 4.7\% and 5.2\% for light mesons and baryons, respectively. As a consequence,  the associated particle yield from the di-hadron correlation is predicted to be enhanced by a factor of $\sim25$ at this \pT range. The enhancement could be even larger if we take into account the ``blue shift'' of \pT spectra induced by radial flow.

In summary, we have applied the Tsallis Blast-Wave (TBW) model to all the identified particle spectra in \pp collisions at $\sqrt{s}=200$, 540, 900, 7000~GeV. The TBW function fits the data quite well over a broad transverse momentum range (0-10 GeV/$c$). The average radial flow velocity extracted from the fit is consistent with zero in \pp collisions at $\sqrt{s}=200$, 540 GeV and increases to $0.264\pm0.005$ at $\sqrt{s}=900$~GeV and $0.320\pm0.005$ at 7 TeV. We have also tested the \mT scaling behavior of the particle spectra. The particle spectra was found to obey \mT scaling at 200~GeV and 540~GeV, but significantly deviate from \mT scaling at beam energy above 900~GeV. The breaking of the \mT scaling at high-energy \pp collisions may be attributed to radial flow. This is suggestive of an onset of radial flow at certain beam energy where sufficient energy density could generate collective motion to be observed in minimum bias \pp collisions.

\begin{acknowledgments}
This work was supported in part by the National Nature Science Foundation of China under Grant Nos 1100504, 11375172 and 1100503, the Offices of NP and HEP within the U.S. DOE Office of Science under the contract of DE-AC02-98CH10886.
\end{acknowledgments}


\bibliography{TBW_LHC_pp_v8}

\end{document}